\documentclass[10pt,english,prd,superscriptaddress,nofootinbib,preprintnumbers,twocolumn, showpacs,floatfix]{revtex4-2}

\usepackage[utf8]{inputenc}
\usepackage{amsmath}
\usepackage{amsfonts}
\usepackage{amssymb}
\usepackage{graphicx} 
\usepackage[caption=false]{subfig}
\graphicspath{ {figures/} }
\usepackage[usenames,dvipsnames]{xcolor}
\usepackage{hyperref}   
\hypersetup{
	colorlinks=true, 
	citecolor=MidnightBlue,
	linkcolor=MidnightBlue,
	urlcolor=Cyan}
\usepackage{tikz}
\usepackage{verbatim}
\usepackage{soul}
\definecolor{purple}{rgb}{1,0,1}
\definecolor{lime}{HTML}{A6CE39}

\def\be{\begin{equation}}
	\def\ee{\end{equation}}
\def\bea{\begin{eqnarray}}
	\def\eea{\end{eqnarray}}

\begin{document}
	
\title{Exact general relativistic solutions for a cylindrically symmetric stiff fluid matter source}  

\author{Tiberiu Harko}
\email{tiberiu.harko@aira.astro.ro}
\affiliation{Department of Physics, Babe\c s-Bolyai University, Kog\u alniceanu Street, Cluj-Napoca, 400084, Romania}
\affiliation{Astronomical Institute of the Romanian Academy, Cluj-Napoca Branch, 19 Cire\c silor Street, 400487 Cluj-Napoca, Romania,}
\author{Francisco S. N. Lobo}
\email{fslobo@ciencias.ulisboa.pt}
\affiliation{Instituto de Astrof\'{i}sica e Ci\^{e}ncias do Espa\c{c}o, Faculdade de Ci\^{e}ncias da Universidade de Lisboa, Edifício C8, Campo Grande, P-1749-016 Lisbon, Portugal}
\affiliation{Departamento de F\'{i}sica, Faculdade de Ci\^{e}ncias da Universidade de Lisboa, Edif\'{i}cio C8, Campo Grande, P-1749-016 Lisbon, Portugal}
\author{Man Kwong Mak}
\email{mankwongmak@gmail.com}
\affiliation{Departamento de F\'{i}sica, Facultad de Ciencias Naturales, Universidad de Atacama, Copayapu 485, Copiap\'o, Chile.}

\date{\LaTeX-ed \today}

\begin{abstract}
	
	In this work, we derive the general solutions for a cylindrically symmetric space-time filled with a cosmological perfect fluid obeying $p=\gamma \rho$ ($0\leq \gamma \leq 1$), where $\gamma=1$ represents a stiff or Zeldovich fluid. Using Marder's metric with coefficients depending on $t$ and $r$, we obtain explicit solutions of the gravitational field equations for the three cases $\delta = 1, 0, -1$, corresponding to exponential, power-law, and trigonometric behaviors of the metric functions. The resulting space-times exhibit anisotropic evolution, nontrivial expansion and shear, and curvature singularities, with energy density and pressure profiles determined by the integration constants. These solutions provide a comprehensive framework for modeling cylindrically symmetric cosmologies, offering insights into early-universe dynamics and anisotropic gravitational phenomena. The versatility of the solutions also opens avenues for extensions to higher-dimensional or modified gravity scenarios, making them a valuable tool for both theoretical and phenomenological studies in general relativity.
	
\end{abstract}

\maketitle

\tableofcontents

\section{Introduction}

The standard Friedmann--Lema\^{i}tre--Robertson--Walker (FLRW) cosmological model, based on the assumption of a homogeneous and isotropic distribution of matter and radiation, together with the existence of a hot big-bang origin, has been remarkably successful in describing the large-scale evolution of the Universe. In particular, it provides an accurate framework for explaining several key observational phenomena such as the Hubble expansion, the primordial abundance of light elements predicted by big-bang nucleosynthesis, and the existence of the cosmic microwave background radiation (CMBR). Within this framework, the currently accepted theories of structure formation, especially inflationary cosmology combined with the cold dark matter paradigm, describe the emergence of galaxies and large-scale structures through the gravitational instability of small primordial perturbations present in the early Universe.

Despite these successes, it has become increasingly clear that the assumption of perfect homogeneity and isotropy cannot be strictly valid at all spatial and temporal scales. In particular, observations indicate that the early Universe contained small but significant deviations from perfect uniformity. Measurements of the CMBR by the COBE satellite \cite{COBE1,COBE2,COBE3} revealed tiny temperature anisotropies, which strongly suggest that the primordial Universe was slightly inhomogeneous and anisotropic. These small fluctuations are believed to have acted as the seeds that later evolved gravitationally into the complex structures observed today. Additional observational evidence supporting departures from perfect homogeneity has been obtained from galaxy redshift surveys \cite{LGH,SAUN} and from infrared surveys such as the Infrared Astronomical Satellite (IRAS) survey, which indicate that matter in the Universe is distributed in a highly non-trivial manner on a wide range of scales. These findings, together with the theoretical motivation to avoid extremely special initial conditions for the Universe, have stimulated renewed interest in the study of inhomogeneous cosmological models.

Indeed, recent high-precision observations of the CMBR provide increasingly stringent tests of the fundamental predictions of inflation regarding the nature and spectrum of primordial fluctuations \cite{1,2,3,4,5}. Furthermore, observational studies of the large-scale distribution of matter in the Universe have raised questions regarding the exact validity of the cosmological principles of homogeneity and isotropy \cite{an}. In fact, the most recent results obtained from the Planck mission suggest that the presence of a certain degree of intrinsic large-scale anisotropy in the Universe cannot be excluded {\it a priori}. These considerations make the study of cosmological models with reduced symmetry particularly relevant.

	Cylindrically symmetric spacetimes arise naturally in several physically relevant contexts beyond their mathematical interest as a simplified anisotropic geometry. 
	Cosmic strings, which are one-dimensional topological defects that may have formed during phase transitions in the early Universe, are predicted by a number of grand-unified theories and can act as seeds for large-scale structure formation~\cite{Vilenkin:1984ib,Hindmarsh:1994re}.
	The spacetime surrounding an isolated straight cosmic string is approximately cylindrically symmetric, and exact solutions provide useful approximations for studying the gravitational field of such objects \cite{Harko2015a,Harko2015b,Harko2020,Harko2021}.
	More generally, cylindrical models serve as prototypes for anisotropic cosmological dynamics, capturing essential features of shear and inhomogeneity that are absent in the FLRW framework yet may be relevant during the early phases of cosmic evolution~\cite{Bicak:2000hh}.
	Recent work has also explored cylindrically symmetric geometries in the context of gravitational collapse, gravitational waves with extended sources, and as toy models for testing modified gravity theories~\cite{Bronnikov:2020cyl}.

Among the various classes of anisotropic and inhomogeneous space-times, cylindrically symmetric models occupy a significant position in relativistic cosmology. Such models provide valuable insights into gravitational phenomena where anisotropy and spatial inhomogeneity cannot be neglected. The study of cylindrically symmetric solutions in general relativity dates back to the pioneering works of Levi-Civita and Weal in 1917--1918, and since then an extensive body of literature has been devoted to their investigation. In particular, static fluid cylinders have been widely studied \cite{EVAN,BRON,SANTOS,TYAGI}, leading to the construction of large classes of exact interior and exterior solutions of Einstein's field equations for cylindrically symmetric configurations. These investigations include models for fluids obeying different equations of state and also consider the possible presence of electromagnetic fields. A family of exact solutions to the Einstein--Maxwell equations for rotating cylindrically symmetric distributions of a perfect fluid satisfying the equation of state $p = w\rho$, $|w|<1$, carrying a circular electric current in the angular direction was obtained in \cite{Bron1}. Cylindrically symmetric solutions the  $f(R,\varphi,X)$ theory of gravity were considered in \cite{Adnan}.

Of special interest in relativistic astrophysics and cosmology is the problem of the gravitational collapse of a self-gravitating fluid, especially in the case where the pressure equals the energy density, corresponding to a stiff fluid equation of state. For cylindrically symmetric space-times, several exact stiff-fluid solutions with metric coefficients depending on the variables $t$ and $r$ have been obtained in \cite{TAUB,LETE1,LETE2,TABEN}. Furthermore, a one-parameter family of solutions describing the gravitational collapse of a stiff fluid was constructed in \cite{DAVID}. The late stages of gravitational collapse, as well as the interaction and collision of cylindrically symmetric null fluids, have also been investigated through analytical models \cite{WANG}. Static cylindrically symmetric solutions of Einstein's field equations including a cosmological constant, which can be interpreted as describing cosmic strings, have also been explored \cite{Linet:1986sr}.

All static, cylindrically symmetric solutions of the Einstein field equations in vacuum were obtained in \cite{Tref}, including not only the locally flat solutions, but also others in which the metric coefficients are powers of the radial coordinate, and the spacetime is curved. Since all the vacuum solutions are singular on the axis, a matching with  interior solutions with nonvanishing energy density and pressure was also performed. The mechanical stability of a static cylindrically symmetric regular solution to Einstein's field equations, supported by a static uniform perfect fluid with the equation of state $p = -\rho/3$  was considered in \cite{Maz}.

One of the most striking predictions of relativistic cosmology is the existence of an initial singularity, commonly referred to as the big-bang singularity. The occurrence of such a singularity follows from the well-known singularity theorems of general relativity, which show that under physically reasonable conditions, such as the positivity of the energy density, causality requirements, and regularity assumptions, an initial cosmological singularity is unavoidable if Einstein's field equations remain valid. Avoiding such singularities generally requires the inclusion of quantum gravitational effects or modifications of classical general relativity. 

In this context, Senovilla and collaborators \cite{SEN,FLGH} obtained an important class of exact solutions of Einstein's equations describing a singularity-free cosmological model. These solutions represent cylindrically symmetric universes filled with a perfect fluid, with the metric tensor components depending on the variables $t$ and $z$. Remarkably, these models are completely regular everywhere and satisfy the standard energy and causality conditions. Subsequently, a broader class of singularity-free cosmological models was obtained through a detailed investigation of cylindrically symmetric metrics with separable functions of $r$ and $t$ as metric coefficients \cite{RSEN}. Earlier related work by Wainwright and Goode \cite{GOOD} had already produced a class of cosmological models with two spacelike commuting Killing vectors, known as orthogonally transitive cosmologies, which exhibit interesting dynamical behaviors and do not necessarily approach spatial homogeneity either at late times or near singularities.

Senovilla and Vera also derived an explicit exact solution of Einstein's equations describing an inhomogeneous dust-filled universe with cylindrical symmetry, which possesses several intriguing physical properties \cite{Senovilla:2000cr}. In this model the universe is effectively ``closed'': the dust expands from a big-bang singularity and eventually recollapses to a big-crunch singularity. Interestingly, both singularities are connected, so that the entire space-time is enclosed within a single singular hypersurface of general character. Moreover, the big-bang is not simultaneous for all observers; the age of the Universe as measured by dust particles depends on their spatial position due to the intrinsic inhomogeneity of the model, and the total lifetime of the dust worldlines has no non-zero lower bound. For the physical and astrophysical implications of the static solutions of the Einstein equations in cylindrical symmetry see  \cite{N1, N2,N3,N4,N5,N6,N7}. Some special solutions to the Einstein-Maxwell equations  corresponding to source-free steady-state cylindrically symmetric electromagnetic fields have been obtained in \cite{Santos1}.

Inhomogeneous cosmological models containing scalar fields have also been investigated within the context of inflationary cosmology. Such models were studied in \cite{IBAN} in order to analyze how inflation and isotropization depend on the degree of space-time inhomogeneity. These investigations were further generalized in \cite{LABRA,FIBAN}, where a broad class of inhomogeneous solutions containing scalar fields with exponential potentials was constructed, extending the spatially flat FLRW cosmologies. An intriguing feature of these inhomogeneous scalar-field models is that they generally fail to produce inflation and do not necessarily evolve towards isotropy. Additionally, the influence of dissipative processes, such as bulk viscosity and heat flux, as well as the presence of electromagnetic fields in cylindrically symmetric inhomogeneous cosmological models has also been explored \cite{KILL}. For recent reviews of the role of the cylindrically symmetric fields in general relativity see \cite{Bron} and \cite{Santos}, respectively. 

The purpose of the present paper is to investigate a class of cosmologically relevant solutions for a cylindrically symmetric space-time filled with a cosmological fluid obeying a Zeldovich equation of state, also known as a stiff fluid equation of state. Such an equation of state, characterized by the relation $p=\rho$, represents the limiting case of a relativistic perfect fluid and has been widely considered in the context of early-universe cosmology and scalar-field dynamics. The space-time metric is assumed to have the form introduced by Marder \cite{MARDER}, with the metric tensor coefficients depending on the variables $t$ and $r$, thus allowing for both temporal evolution and radial inhomogeneity. Within this framework, we derive the general solutions of the gravitational field equations and analyze their main geometrical and physical properties. In particular, we examine the structure of the resulting space-times and discuss the role played by the integration constants in determining the qualitative behaviour of the cosmological models.

This paper is organized as follows: In Sec.~\ref{SecII}, we present the Einstein field equations corresponding to the adopted cylindrically symmetric metric. In Sec.~\ref{SecIII}, the general solution of the field equations for a cosmological fluid obeying the Zeldovich equation of state is obtained. In Sec.~\ref{SecIV} we discuss the main physical properties of the resulting solutions. Finally, in Sec.~\ref{SecV} we summarize our results and present the main conclusions of this work.

\section{Geometry and gravitational field equations}\label{SecII}

The most general line element describing a space-time with cylindrical symmetry with respect to the $z$ axis may be written in the form
\begin{equation}
	ds^{2}=e^{2F_{0}}dt^{2}-e^{2F_{1}}dr^{2}-e^{2F_{2}}d\phi
	^{2}-e^{2F_{3}}dz^{2} \,, \label{1a}
\end{equation}
where the functions $F_{i}=F_{i}(t,r)$, with $i=0,1,2,3$, depend in general on both the time coordinate $t$ and the radial coordinate $r$. This metric represents the most general cylindrically symmetric geometry with two commuting Killing vectors associated with the coordinates $\phi$ and $z$ \cite{MARDER}. In the present work we use the natural system of units with $8\pi G=c=1$.

We assume that the spacetime is filled with a cosmological fluid characterised by an energy density $\rho$ and an isotropic pressure $p$. The corresponding energy-momentum tensor for a perfect fluid is therefore given by
\begin{equation}
	T_{0}^{0}=\rho , \qquad T_{1}^{1}=T_{2}^{2}=T_{3}^{3}=-p \,, \label{7a}
\end{equation}
where the remaining components vanish. We further assume that the fluid
obeys a linear barotropic equation of state
\[
p=\gamma \rho , \qquad 0\leq \gamma \leq 1 ,
\]
where the parameter $\gamma$ determines the physical nature of the fluid. For instance, $\gamma=0$ corresponds to dust, $\gamma=1/3$ to radiation, while $\gamma=1$ represents a stiff (Zeldovich) fluid.
For a comoving fluid configuration the four-velocity is aligned with the temporal coordinate and takes the form $u_{i}=e^{F_{0}}\delta _{i}^{0}$. This choice implies that the matter distribution is at rest in the adopted coordinate system.

In order to simplify the form of the metric, we specialise the line element by imposing the condition $F_{0}=F_{1}$. With this choice the metric reduces to the form introduced by Marder \cite{MARDER}, and the $(t,r)$ sector of the metric becomes conformally flat. In particular, the line element becomes Lorentz invariant in the $(t,r)$ plane, which considerably simplifies the structure of the field equations.

The gravitational dynamics of the system are determined by Einstein's field equations, which can be written as
\begin{equation}
	R_{i}^{k}=T_{i}^{k}-\frac{1}{2}\delta _{i}^{k}T  \,, \label{8a}
\end{equation}
where $R_{i}^{k}$ is the Ricci tensor and $T=T_{i}^{i}$ is the trace of the energy-momentum tensor.

\paragraph{Field equations.} For the metric (\ref{1a}) with the condition $F_{0}=F_{1}$, the field equations reduce, after some algebraic manipulation, to the following set of partial differential equations
\begin{equation}
	\frac{\partial \left( \ln u\right) }{\partial r}\frac{\partial 
		F_{1} }{\partial t}+\frac{\partial \left( \ln u\right) }{%
		\partial t}\frac{\partial F_{0} }{\partial r}-
	\left[ \frac{\partial ^{2}\left( \ln u\right) }{\partial t\partial r}%
	+h_{2}k_{2}+h_{3}k_{3}\right] =0  \,,   \label{EF1}
\end{equation}
\begin{eqnarray}
	3\frac{\partial 
		H}{\partial t}+3h_{0}H+3\bar{H}^{2}&-&\frac{1}{\Omega }\frac{\partial \left( \Omega
		K_{0}\right) }{\partial r}\nonumber\\
&&=-\frac{1}{2}\left( 1+3\gamma \right) \rho  \,,
	\label{EF2}
\end{eqnarray}
\begin{eqnarray}
	-\frac{1}{\Omega }\frac{\partial \left( \Omega H_{1}\right) }{\partial t}-%
	\frac{1}{\Omega }\frac{\partial \left( \Omega K_{1}\right) }{\partial r}&+&4%
	\frac{\partial K}{\partial r}+4\bar{K}^{2}-2\bar{k}_{1}^{2}+8k K_1 \nonumber\\
&&=-\frac{1}{2}%
	\left( 1-\gamma \right) \rho  \,,  \label{EF3}
\end{eqnarray}
\begin{equation}
	-\frac{1}{\Omega }\frac{\partial \left( \Omega H_{i}\right) }{\partial t}+%
	\frac{1}{\Omega }\frac{\partial \left( \Omega K_{i}\right) }{\partial r}=-%
	\frac{1}{2}\left( 1-\gamma \right) \rho \,, \qquad i=2,3  \,. \label{EF4}
\end{equation}

For notational simplicity, we have introduced the following auxiliary quantities
\begin{eqnarray}
	h_{i}&=&\dot{F}_{i}, \qquad  H_{i}=e^{-2F_{0}}h_{i}, \;\;  i=0,1,2,3,  \label{2a}
	\nonumber	\\
	h&=&\frac{1}{3}\left( h_{1}+h_{2}+h_{3}\right), \qquad  H=e^{-2F_{0}}h, \nonumber\\
	 \bar{h}^{2}&=&
	\frac{1}{3}\left( h_{1}^{2}+h_{2}^{2}+h_{3}^{2}\right) , \qquad  \bar{H}%
	^{2}=e^{-2F_{0}}\bar{h}^{2},  \label{3a}
	\nonumber	\\
	k_{i}&=&F_{i}^{\prime },  \;\;   K_{i}=e^{-2F_{1}}k_{i},   
	\bar{k}_{i}^{2}=e^{-2F_{1}}k_{i}^{2},  \;\;  i=0,1,2,3,  \label{4a}
	\nonumber	\\
	k&=&\frac{1}{4}\left( k_{0}+k_{1}+k_{2}+k_{3}\right),  \qquad   K=e^{-2F_{1}}k, \nonumber\\
	\bar{k}^{2}&=&\frac{1}{4}\left( k_{0}^{2}+k_{1}^{2}+k_{2}^{2}+k_{3}^{2}\right),  \qquad 
	\bar{K}^{2}=e^{-2F_{1}}\bar{k}^{2}  \label{5a}
	\nonumber	\\
	\Omega &=&e^{F_{0}+F_{1}+F_{2}+F_{3}},  \qquad  u=e^{F_{2}+F_{3}},  
	\nonumber	\label{6a}
\end{eqnarray}
where a dot and a prime denote partial differentiation with respect to the time coordinate $t$ and the radial coordinate $r$, respectively. The quantities $h_{i}$ and $k_{i}$, $i=1,2,3$ therefore represent the temporal and radial derivatives of the metric functions, respectively.

\paragraph{Conservation equations.} The conservation of the energy-momentum tensor, $T^{k}{}_{i;k}=0$, provides additional relations between the dynamical variables. Taking into account the equation of state $p=\gamma \rho$, the components $T^{k}{}_{0;k}=0$ and $T^{k}{}_{1;k}=0$ lead to the conservation equations
\begin{equation}
	\dot{\rho}+3\left( 1+\gamma \right) h\rho =0, \qquad  p^{\prime }+k_{0}\left(
	1+\gamma \right) \rho =0 \,,  \label{9a}
\end{equation}
which describe the temporal and spatial evolution of the energy density of the fluid.

By adding Eqs.~(\ref{EF4}) for the appropriate components, one obtains the following second-order differential equation for the function $u$
\begin{equation}
	\frac{\partial ^{2}u}{\partial t^{2}}-\frac{\partial ^{2}u}{\partial r^{2}}%
	=\left( 1-\gamma \right) \rho e^{2F_{0}}u \,. \label{10a}
\end{equation}
This equation plays a central role in the determination of the metric functions.

It is also useful to rewrite Eqs.~(\ref{EF4}) in the equivalent form
\begin{equation}
	\frac{\partial \left( u\dot{F}_{i}\right) }{\partial t}-\frac{\partial
		\left( uF_{i}^{^{\prime }}\right) }{\partial r}=\frac{1}{2}\left( 1-\gamma
	\right) \rho e^{2F_{0}}u, \qquad  i=2,3.  \label{11a}
\end{equation}
Subtracting Eq. (\ref{10a}) from Eq. (\ref{11a}) and introducing the new variables
\[
\nu _{i}=F_{i}-\frac{1}{2}\ln u, \qquad i=2,3 ,
\]
we obtain the simplified relation
\begin{equation}
	\frac{\partial \left( u\dot{\nu}_{i}\right) }{\partial t}-\frac{\partial
		\left( u\nu _{i}^{{\prime }}\right) }{\partial r}=0, \qquad   i=2,3 \,.  \label{12a}
\end{equation}
This form considerably simplifies the analysis of the metric functions associated with the angular and axial directions.

\paragraph{The case of the linear barotropic equation of state.} In the case of a fluid obeying the equation of state $p=\gamma  \rho$, $\gamma ={\rm constant}$, the Bianchi identities (\ref{9a}) can be integrated directly. Performing the integration yields
\begin{equation}
	e^{F_{0}}=f^{-1}\left( t\right) \rho ^{-\frac{\gamma }{1+\gamma }%
	} , \;\;   u=e^{F_{2}+F_{3}}=f\left( t\right) g\left( r\right) \rho ^{-\frac{1-\gamma
		}{1+\gamma }},  \label{13a}
\end{equation}
where $f(t)$ and $g(r)$ are arbitrary functions arising from the
integration.

Substituting the expressions (\ref{13a}) into Eq. (\ref{10a}), the latter reduces to the simpler, hyperbolic type inhomogeneous second order partial differential equation
\begin{equation}
	\frac{\partial ^{2}u}{\partial t^{2}}-\frac{\partial ^{2}u}{\partial r^{2}}%
	=\left( 1-\gamma \right) \frac{g\left( r\right) }{f\left( t\right) }\,.
	\label{14a}
\end{equation}

Thus, the fundamental equations governing the dynamics of a cylindrically symmetric fluid distribution obeying the equation of state $p=\gamma \rho$ have been obtained. In particular, the determination of the metric coefficients $e^{F_{2}}$ and $e^{F_{3}}$ requires the solution of the coupled equations (\ref{14a}) and (\ref{12a}). Once the function $u$ is
known, the coefficient $e^{F_{0}}$ can be obtained either by solving Eq. (\ref{EF1}) or directly from the relation
\[
e^{F_{0}}=\rho (t)^{-\frac{1}{1-\gamma }}
\left[ \frac{u}{g(r)} \right] ^{\frac{\gamma }{1-\gamma }} .
\]
The energy density $\rho$ itself follows from the second conservation equation in (\ref{9a}).

In the following sections we shall determine explicit solutions of the gravitational field equations for the case of a stiff fluid, corresponding to $\gamma=1$.

\section{General solutions of the gravitational field equations for a stiff fluid}\label{SecIII}

In this section, we determine the general solutions of the gravitational field equations obtained in the previous section for the particular case of a stiff fluid. A stiff fluid corresponds to the equation of state $p=\rho$, that is, $\gamma =1$. This equation of state is of particular interest in
relativistic cosmology since it represents the limiting case in which the speed of sound in the fluid equals the speed of light. Such a fluid, also known as a Zeldovich fluid, may play an important role in the description of matter at extremely high densities, such as those that might occur in the very early Universe.

\paragraph{General solution for a stiff fluid.} For $\gamma =1$, the equations describing the dynamics of the fluid simplify considerably. In particular, Eqs. (\ref{13a}) and (\ref{14a}) reduce to
\begin{equation}
	u=f\left( t\right) g\left( r\right), \qquad  \frac{\ddot{f}}{f}=\frac{g^{\prime
			\prime }}{g}=\delta \lambda ^{2} \,, \label{15a}
\end{equation}
where $\delta =0,\pm 1$ and $\lambda$ is a separation constant. The above relation follows from the fact that the function $u$ becomes separable in the variables $t$ and $r$. Consequently, the differential equation for $u$ can be separated into two ordinary differential equations for the functions $f(t)$ and $g(r)$, respectively.

The parameter $\delta$ determines the qualitative behavior of the solutions and leads to three distinct classes of solutions depending on whether the separation constant is positive, zero, or negative. Solving the differential equations for $f(t)$ and $g(r)$ yields the following general expressions for the function $u$
\begin{widetext}
\begin{equation}
	u=\left\{
	\begin{array}
		[c]{l}%
		\left( c_{1}e^{\lambda t}+c_{2}e^{-\lambda t}\right) \left(
		c_{3}e^{\lambda r}+c_{4}e^{-\lambda r}\right), \qquad  \delta =1  \,\\
		\\
		\left( at+b\right) \left( cr+d\right) \,,   \qquad  \delta =0    \\
		\\
		\left[ d_{1}\cos \left( \lambda t\right) +d_{2}\sin \left( \lambda
		t\right) \right] \left[ d_{3}\cos \left( \lambda r\right) +d_{4}\sin \left(
		\lambda r\right) \right], \qquad \delta =-1
	\end{array}
	\right.   \label{16a}
\end{equation}
\end{widetext}
where $c_{i}$, $d_{i}$ (and $a$, $b$, $c$, $d$) are integration constants, with $i=1,2,3,4$. These three classes correspond respectively to exponential, linear, and oscillatory behaviors of the metric function $u$.

For the stiff fluid equation of state, Eq. (\ref{11a}) simplifies considerably since the term proportional to $(1-\gamma)$ vanishes. As a result, Eq. (\ref{11a}) reduces to
\begin{equation}
	\frac{\partial \left( u\dot{F}_{i}\right) }{\partial t}-\frac{\partial
		\left( uF_{i}^{^{\prime }}\right) }{\partial r}=0, \qquad  i=2,3.  \label{17a}
\end{equation}
This equation governs the behavior of the metric functions $F_{2}$ and $F_{3}$ associated with the angular and axial coordinates.

\paragraph{The separability condition.} In order to simplify the integration of Eq. (\ref{17a}) and to ensure consistency with the relation $\ln u=F_{2}+F_{3}$, we assume that the metric functions can be written in a separable form
\[
F_{i}\left( t,r\right) =A_{i}\left( t\right) +B_{i}\left( r\right),
\qquad i=2,3 .
\]
This assumption allows us to reduce the partial differential equations into ordinary differential equations for the functions $A_{i}(t)$ and $B_{i}(r)$.

For notational convenience, and in order to simplify the expressions appearing in the subsequent calculations, we introduce the auxiliary functions
\begin{eqnarray}
	\hspace{-0.5cm}\tau &=&\frac{1}{\lambda }\left( c_{1}e^{\lambda t}-c_{2}e^{-\lambda t}\right),
	  \eta =\frac{1}{\lambda }\left( c_{3}e^{\lambda r}-c_{4}e^{-\lambda
		r}\right) ,  \label{18a}
	\\
	\hspace{-0.5cm}\sigma &=&\frac{1}{\lambda }\left[ d_{1}\sin \left( \lambda t\right) -d_{2}\cos
	\left( \lambda t\right) \right], \\
	\hspace{-0.5cm}\theta& =&\frac{1}{\lambda }\left[ d_{3}\sin
	\left( \lambda r\right) -d_{4}\cos \left( \lambda r\right) \right] \,,
	\label{19a}
\end{eqnarray}
which are constructed so as to simplify the representation of the
solutions for the exponential and oscillatory cases.

These auxiliary variables will prove useful in expressing the metric functions in a compact form and in facilitating the integration of the field equations. The general solutions will therefore be analyzed below by considering separately the three possible cases corresponding to $\delta =1$, $\delta =0$, and $\delta =-1$.

\subsection{Specific case: $\delta =1$}

We first consider the case $\delta =1$, which corresponds to the
exponential class of solutions obtained in Eq. (\ref{16a}). In this case the function $u=f(t)g(r)$ has the form
\[
u=\left( c_{1}e^{\lambda t}+c_{2}e^{-\lambda t}\right)
\left( c_{3}e^{\lambda r}+c_{4}e^{-\lambda r}\right),
\]
and the auxiliary variables $\tau$ and $\eta$, introduced in
Eqs.~(\ref{18a})--(\ref{19a}), provide a convenient parametrization of the solutions.

\paragraph{The metric tensor components $F_2$ and $F_3$.} Substituting the above expressions into Eq. (\ref{17a}), and integrating the resulting differential equations, the general solution for the metric coefficients $F_{2}$ and $F_{3}$ can be written as
\begin{widetext}
\begin{equation}
	e^{2F_{i}}=\left( \lambda ^{2}\tau ^{2}+4c_{1}c_{2}\right) ^{\frac{\alpha
			_{i}}{\lambda ^{2}}}\left( \lambda ^{2}\eta ^{2}+4c_{3}c_{4}\right) ^{\frac{%
			\alpha _{i}}{\lambda ^{2}}}e^{\frac{2Q_{i}}{\lambda ^{2}}F\left( \tau
		\right) +\frac{2R_{i}}{\lambda ^{2}}G\left( \eta \right) }, \;\; i=2,3,
	\label{20}
\end{equation}
\end{widetext}
where $\alpha_{i}$ are separation constants and $Q_{i}$ and $R_{i}$ are integration constants. For notational convenience, we have introduced the functions
\begin{equation}
	F\left( \tau \right) =\int \frac{d\tau }{\tau ^{2}+4\lambda
		^{-2}c_{1}c_{2}}, \;\;  G\left( \eta \right) =\int \frac{d\eta }{\eta ^{2}+4\lambda
		^{-2}c_{3}c_{4}} \,,
\end{equation}
which arise naturally in the integration of Eq. (\ref{17a}).

The integral defining $F(\tau)$ can be evaluated explicitly. Depending on the sign of the product $c_{1}c_{2}$, one obtains
\begin{equation}
	F(\tau)=\left\{
	\begin{array}
		[c]{l}%
		n\tan ^{-1}\left( n\tau \right), \qquad  c_{1}c_{2}>0 \,,\\
		\\
		\frac{n}{2}\ln \left| \frac{n\tau -1}{n\tau +1}\right|, \qquad  c_{1}c_{2}<0 \,,
	\end{array}
	\right.  \label{20a}
\end{equation}
where $n=\frac{\lambda }{2\sqrt{c_{1}c_{2}}}$. The function $G(\eta)$ is
obtained in an analogous way and depends on the sign of the product $c_{3}c_{4}$.

The constants appearing in Eq. (\ref{20}) are not independent. In fact, they must satisfy the relations
\begin{equation}
	\alpha _{1}+\alpha _{2}=\lambda ^{2}, \qquad   Q_{2}+Q_{3}=0,  \qquad R_{2}+R_{3}=0 \,. \label{22}
\end{equation}

These constraints follow directly from the condition $\ln u=F_{2}+F_{3}$, which relates the metric functions through the definition of the variable $u$.

\paragraph{The metric component $F_0$.} In order to determine the remaining metric function $F_{0}$, we substitute the expressions for $F_{2}$ and $F_{3}$ into Eq.~(\ref{EF1}). This leads to the following first-order linear partial differential equation for $F_{0}$
\begin{widetext}
\begin{eqnarray}
	&&\frac{\lambda ^{2}\left( \lambda ^{2}\tau ^{2}+4c_{1}c_{2}\right) }{\tau }%
	\frac{\partial F_{0}}{\partial \tau }+\frac{\lambda ^{2}\left( \lambda
		^{2}\eta ^{2}+4c_{3}c_{4}\right) }{\eta }\frac{\partial F_{0}}{\partial \eta
	}-2\left( \alpha _{2}^{2}+\alpha _{3}^{2}\right) 
	\nonumber  \\
	&&-\frac{2\left( \alpha _{1}Q_{2}+\alpha _{3}Q_{3}\right) }{\tau }-\frac{%
		2\left( \alpha _{2}R_{2}+\alpha _{3}R_{3}\right) }{\eta }-\frac{2\left(
		Q_{2}R_{2}+Q_{3}R_{3}\right) }{\tau \eta }=0 \,, \label{25}
\end{eqnarray}
\end{widetext}
whose general solution can be written as
\begin{widetext}
\begin{equation}
	e^{2F_{0}}=\left[ \left( \lambda ^{2}\tau ^{2}+4c_{1}c_{2}\right) \left(
	\lambda ^{2}\eta ^{2}+4c_{3}c_{4}\right) \right] ^{q_{1}}e^{q_{2}F\left(
		\tau \right) +q_{3}G\left( \eta \right) +q_{4}L\left( x\right) }\Phi \left(
	\ln \left| \frac{\lambda ^{2}\tau ^{2}+4c_{1}c_{2}}{\lambda ^{2}\eta
		^{2}+4c_{3}c_{4}}\right| \right) ,  \label{27}
\end{equation}
\end{widetext}
where $\Phi$ is an arbitrary function and
\[
x=\lambda ^{2}\tau \eta -\left[
\left( \lambda ^{2}\tau ^{2}+4c_{1}c_{2}\right)
\left( \lambda ^{2}\eta ^{2}+4c_{3}c_{4}\right)
\right]^{1/2}.
\]

The function $L(x)$ is defined by the integral
\[
L\left( x\right)=\int \frac{dx}{x^{2}-16m^{2}}, \qquad
m=\sqrt{c_{1}c_{2}c_{3}c_{4}},
\]
which yields the explicit expressions
\begin{equation}
	L(x)=\left\{
	\begin{array}
		[c]{l}%
		\frac{1}{4m}\tan ^{-1}\frac{x}{4m},  \qquad  m<0 \,,\\
		\\
		\frac{1}{8m}\ln \left| \frac{x-4m}{x+4m}\right|, \qquad  m>0 \,.
	\end{array}
	\right.   \label{20aa}
\end{equation}

The constants $q_{i}$, for $i=1,2,3,4$, appearing in Eq. (\ref{27}) are
given by
\begin{eqnarray}\label{29}
	q_{1}&=&\lambda ^{-4}\left( \alpha _{2}^{2}+\alpha _{3}^{2}\right)
	, \qquad \;\;\; q_{2}=4\lambda ^{-4}\left( \alpha _{2}Q_{2}+\alpha _{3}Q_{3}\right), \nonumber\\
	   q_{3}&=&4\lambda ^{-4}\left( \alpha _{2}R_{2}+\alpha _{3}R_{3}\right)
	,  \; q_{4}=4\lambda ^{-6}\left( Q_{2}R_{2}+Q_{3}R_{3}\right) \,.\nonumber\\
\end{eqnarray}


\paragraph{Particular solutions-vanishing integration constants.} A number of particular solutions can be obtained by choosing specific values of the integration constants appearing in Eq. (\ref{20}). For example, if we set $Q_{i}=R_{i}=0$ ($i=2,3$), the metric functions simplify considerably and the solutions of the field equations take the form
\begin{widetext}
\begin{equation}
	\left\{
	\begin{array}
		[c]{l}%
		e^{2F_{0}}=\left[ \left( c_{1}e^{\lambda t}+c_{2}e^{-\lambda t}\right)
		\left( c_{3}e^{\lambda r}+c_{4}e^{-\lambda r}\right) \right] ^{\frac{\alpha
				_{2}^{2}+\alpha _{3}^{2}}{\lambda ^{4}}}\Sigma \left( \ln \left| \frac{%
			c_{1}e^{\lambda t}+c_{2}e^{-\lambda t}}{c_{3}e^{\lambda r}+c_{4}e^{-\lambda
				r}}\right| \right)  \,,\\
		\\
		e^{2F_{i}}=\left[ \left( c_{1}e^{\lambda t}+c_{2}e^{-\lambda t}\right)
		\left( c_{3}e^{\lambda r}+c_{4}e^{-\lambda r}\right) \right] ^{\frac{2\alpha
				_{i}}{\lambda ^{4}}}, \qquad \qquad i=2,3
	\end{array} 
	\right.    \label{35}
\end{equation}
\end{widetext}
where the separation constants satisfy the relation
\[
\alpha _{2}+\alpha _{3}=\lambda ^{2}.
\]

Substituting the expressions (\ref{35}) into the field equations
(\ref{EF2}) and (\ref{EF3}), we obtain a differential equation for the function $\Sigma$:
\begin{equation}
	\left( e^{2x}-\frac{c_{1}c_{2}}{c_{3}c_{4}}\right) \frac{d^{2}\Sigma }{dx^{2}%
	}+2e^{2x}\frac{d\Sigma }{dx}-2e^{2x}=0,  \label{38}
\end{equation}
where
\[
x=\ln \left| \frac{c_{1}e^{\lambda t}+c_{2}e^{-\lambda t}}{%
	c_{3}e^{\lambda r}+c_{4}e^{-\lambda r}}\right|.
\]

Equation~(\ref{38}) admits the general solution
\begin{equation}
	\Sigma \left( x\right) =x+\frac{a_{0}c_{3}c_{4}}{c_{1}c_{2}}\ln \left| \frac{%
		\sqrt{e^{2x}-\frac{c_{1}c_{2}}{c_{3}c_{4}}}}{e^{x}}\right| , 
	\label{40}
\end{equation}
where $a_{0}$ is an integration constant.

Using these results we finally obtain the expressions for the metric coefficient $e^{2F_{0}}$ and for the thermodynamic variables of the fluid, namely the energy density $\rho$ and the pressure $p$
\begin{widetext}
\begin{equation}
	\left\{
	\begin{array}
		[c]{l}%
		e^{2F_{0}}=\left[ \left( \frac{c_{1}e^{\lambda t}+c_{2}e^{-\lambda t}}{%
			c_{3}e^{\lambda r}+c_{4}e^{-\lambda r}}\right) ^{2}-\frac{c_{1}c_{2}}{%
			c_{3}c_{4}}\right] ^{a_{1}}\left( c_{1}e^{\lambda t}+c_{2}e^{-\lambda
			t}\right) ^{a_{2}}\left( c_{3}e^{\lambda r}+c_{4}e^{-\lambda r}\right)
		^{a_{3}}\,,\\
		\\
		p=\rho =\left[
		\left( \frac{c_{1}e^{\lambda t}+c_{2}e^{-\lambda t}}{c_{3}e^{\lambda
				r}+c_{4}e^{-\lambda r}}\right) ^{2}-\frac{c_{1}c_{2}}{c_{3}c_{4}}\right]
		^{-a_{1}}
		\frac{4\left( a_{0}c_{3}c_{4}-2c_{1}c_{2}\right) \lambda ^{2}}{\left( c_{1}e^{\lambda t}+c_{2}e^{-\lambda t}\right)
			^{a_{2}+2}\left( c_{3}e^{\lambda r}+c_{4}e^{-\lambda r}\right) ^{a_{3}}},
	\end{array}  
	\right.    \label{43}
\end{equation}
\end{widetext}
where the constants are
\begin{eqnarray}
	a_{1}&=&\frac{a_{0}c_{3}c_{4}}{2c_{1}c_{2}}, \nonumber\\
	a_{2}&=&\frac{\alpha_{2}^{2}+\alpha _{3}^{2}}{\lambda ^{4}}-\frac{a_{0}c_{3}c_{4}}{c_{1}c_{2}}+1,\nonumber\\
	a_{3}&=&\frac{\alpha _{2}^{2}+\alpha _{3}^{2}}{\lambda ^{4}}+\frac{%
		a_{0}c_{3}c_{4}}{c_{1}c_{2}}-1.\nonumber
\end{eqnarray}

\paragraph{Separable solutions.} Another interesting class of solutions can be obtained by assuming that the metric coefficient $F_{0}$ is separable,
$F_{0}(t,r)=A_{0}(t)+B_{0}(r)$. Under this assumption Eq.~(\ref{EF1}) yields
\begin{equation}
	e^{2F_{0}}=\left( c_{1}e^{\lambda t}+c_{2}e^{-\lambda t}\right) ^{\frac{%
			2\alpha _{0}}{\lambda ^{2}}}\left( c_{3}e^{\lambda r}+c_{4}e^{-\lambda
		r}\right) ^{\frac{2}{\lambda ^{2}}\left( \frac{\alpha _{2}^{2}+\alpha
			_{3}^{2}}{\lambda ^{2}}-\alpha _{0}\right) },   \label{46}
\end{equation}
where $\alpha _{0}$ is a separation constant satisfying
\begin{equation}
	\alpha _{0}=\frac{\alpha _{2}^{2}+\alpha_{3}^{2}+\alpha _{2}\alpha _{3}}{\alpha _{2}+\alpha _{3}}.
\end{equation}

From Eq.~(\ref{EF2}) the corresponding energy density and pressure are obtained as
\begin{equation}
	p=\rho =-\frac{8\left( \alpha _{2}+\alpha _{3}\right) c_{1}c_{2}\left(
		c_{3}e^{\lambda r}+c_{4}e^{-\lambda r}\right) ^{\frac{2\alpha _{2}\alpha _{3}%
			}{\left( \alpha _{2}+\alpha _{3}\right) ^{2}}}}{\left( c_{1}e^{\lambda
			t}+c_{2}e^{-\lambda t}\right) ^{\frac{2\left( \alpha _{0}+\lambda
				^{2}\right) }{\lambda ^{2}}}},   \label{48}
\end{equation}

Therefore, we have obtained a wide class of exact solutions describing a stiff fluid distribution with cylindrical symmetry. Additional particular solutions may be derived by specifying appropriate values for the separation and integration constants appearing in the above expressions.

\subsection{Specific case: $\delta = -1$}

For $\delta = -1$, the general solution of Eq.~(\ref{17a}) takes the form
\begin{widetext}
\begin{equation}
	e^{2F_{i}}=
	(\lambda \sigma -D_{1})^{-\frac{\beta_i}{\lambda^{2}}-\frac{S_i}{D_{1}\lambda}}
	(\lambda \sigma +D_{1})^{-\frac{\beta_i}{\lambda^{2}}+\frac{S_i}{D_{1}\lambda}}
	(\lambda \theta -D_{2})^{-\frac{\beta_i}{\lambda^{2}}-\frac{T_i}{D_{2}\lambda}}
	(\lambda \theta +D_{2})^{-\frac{\beta_i}{\lambda^{2}}+\frac{T_i}{D_{2}\lambda}},
	\label{23}
\end{equation}
\end{widetext}
where $D_{1}^{2}=d_{1}^{2}+d_{2}^{2}$ and $D_{2}^{2}=d_{3}^{2}+d_{4}^{2}$. The separation constants $\beta_i$ and the integration constants $S_i$ and $T_i$, for $i=2,3$, satisfy
\begin{equation}
	\beta_{2}+\beta_{3}+\lambda^{2}=0, 
	\qquad
	S_{2}+S_{3}=0,
	\qquad
	T_{2}+T_{3}=0.
	\label{24}
\end{equation}

\paragraph{The expression of $F_0$.} Proceeding in a manner analogous to the case $\delta = 1$, we obtain the following first-order linear partial differential equation for $F_{0}$:
\begin{widetext}
\begin{eqnarray}
	&&
	\frac{\lambda^{2}(\lambda^{2}\sigma^{2}-D_{1}^{2})}{\sigma}
	\frac{\partial F_{0}}{\partial \sigma}
	+
	\frac{\lambda^{2}(\lambda^{2}\theta^{2}-D_{2}^{2})}{\theta}
	\frac{\partial F_{0}}{\partial \theta}
	-2(\beta_{2}^{2}+\beta_{3}^{2})
	\nonumber
	\\
	&&
	-\frac{2(\beta_{2}S_{2}+\beta_{3}S_{3})}{\sigma}
	-\frac{2(\beta_{2}T_{2}+\beta_{3}T_{3})}{\theta}
	-\frac{2(S_{2}T_{2}+S_{3}T_{3})}{\sigma\theta}
	=0.
	\label{26}
\end{eqnarray}
\end{widetext}

The general solution of Eq.~(\ref{26}) can be written as
\begin{widetext}
\begin{equation}
	e^{2F_{0}}=
	\Psi\!\left(
	\ln \left|
	\frac{\lambda^{2}\sigma^{2}-D_{1}^{2}}
	{\lambda^{2}\theta^{2}-D_{2}^{2}}
	\right|
	\right)
	(\lambda\sigma-D_{1})^{r_{+}}
	(\lambda\sigma+D_{1})^{r_{-}}
	(\lambda\theta-D_{2})^{s_{+}}
	(\lambda\theta+D_{2})^{s_{-}}
	V(\sigma,\theta),
	\label{30}
\end{equation}
\end{widetext}
where $\Psi$ is an arbitrary function of the variables $\sigma $ and $\theta$, and the function $V(\sigma,\theta)$ is defined according to
\begin{equation}
	V(\sigma,\theta)=
	\left[
	\frac{
		\lambda^{2}\sigma\theta
		-
		\sqrt{(\lambda^{2}\sigma^{2}-D_{1}^{2})(\lambda^{2}\theta^{2}-D_{2}^{2})}
		-
		D_{1}D_{2}}
	{
		\lambda^{2}\sigma\theta
		-
		\sqrt{(\lambda^{2}\sigma^{2}-D_{1}^{2})(\lambda^{2}\theta^{2}-D_{2}^{2})}
		+
		D_{1}D_{2}}
	\right]^{r_{5}}.
	\label{31}
\end{equation}

The constants $r_{\pm}$, $s_{\pm}$ and $r_{5}$ are given by
\begin{equation}
	r_{\pm}=
	\frac{1}{\lambda^{3}}
	\left[
	\frac{\beta_{2}^{2}+\beta_{3}^{2}}{\lambda}
	\pm
	\frac{2(\beta_{2}S_{2}+\beta_{3}S_{3})}{D_{1}}
	\right],
\end{equation}
\begin{equation}
	s_{\pm}=
	\frac{1}{\lambda^{3}}
	\left[
	\frac{\beta_{2}^{2}+\beta_{3}^{2}}{\lambda}
	\pm
	\frac{2(\beta_{2}T_{2}+\beta_{3}T_{3})}{D_{2}}
	\right],
\end{equation}
\begin{equation}
	\hspace{-2.5cm}r_{5}=
	\frac{2(S_{2}T_{2}+S_{3}T_{3})}{\lambda^{2}D_{1}D_{2}}.
	\label{32}
\end{equation}

\paragraph{Particular solution for vanishing integration constants.} A particular class of solutions can be obtained by setting $S_i=T_i=0$ for $i=2,3$. In this case the field equations admit the solutions
\begin{widetext}
\begin{equation}
	\left\{
	\begin{array}{l}
		\displaystyle
		e^{2F_{0}}=
		\left\{
		[d_{1}\cos(\lambda t)+d_{2}\sin(\lambda t)]
		[d_{3}\cos(\lambda r)+d_{4}\sin(\lambda r)]
		\right\}^{\frac{2(\beta_{2}^{2}+\beta_{3}^{2})}{\lambda^{4}}}
		\Gamma\!\left[
		\ln\left|
		\frac{d_{1}\cos(\lambda t)+d_{2}\sin(\lambda t)}
		{d_{3}\cos(\lambda r)+d_{4}\sin(\lambda r)}
		\right|
		\right],
		\\
		\\
		\displaystyle
		e^{2F_{i}}=
		\left\{
		[d_{1}\cos(\lambda t)+d_{2}\sin(\lambda t)]
		[d_{3}\cos(\lambda r)+d_{4}\sin(\lambda r)]
		\right\}^{\frac{2\beta_{i}}{\lambda^{2}}},
		\qquad i=2,3 .
	\end{array}
	\right.
	\label{36}
\end{equation}
\end{widetext}

Substituting Eq.~(\ref{36}) into Eqs.~(\ref{EF2}) and (\ref{EF3}) yields the following equation for the function $\Gamma$:
\begin{equation}
	\left(
	e^{2y}-\frac{d_{1}^{2}+d_{2}^{2}}{d_{3}^{2}+d_{4}^{2}}
	\right)
	\frac{d^{2}\Gamma}{dy^{2}}
	+
	2e^{2y}\frac{d\Gamma}{dy}
	-
	2e^{2y}
	=0,
	\label{39}
\end{equation}
where
\[
y=\ln\left|
\frac{d_{1}\cos(\lambda t)+d_{2}\sin(\lambda t)}
{d_{3}\cos(\lambda r)+d_{4}\sin(\lambda r)}
\right|.
\]

The general solution of Eq.~(\ref{39}) is
\begin{equation}
	\Gamma(y)
	=
	y
	+
	\frac{b_{0}(d_{3}^{2}+d_{4}^{2})}{d_{1}^{2}+d_{2}^{2}}
	\ln\left|
	\frac{\sqrt{e^{2y}-\frac{d_{1}^{2}+d_{2}^{2}}{d_{3}^{2}+d_{4}^{2}}}}
	{e^{y}}
	\right|,
	\label{41}
\end{equation}
where $b_{0}$ is an integration constant.

Consequently, the metric function $e^{2F_{0}}$ and the energy-momentum profile become
\begin{widetext}
\begin{equation}
	\left\{
	\begin{array}{l}
		\displaystyle
		e^{2F_{0}}=
		\left[
		\left(
		\frac{d_{1}\cos(\lambda t)+d_{2}\sin(\lambda t)}
		{d_{3}\cos(\lambda r)+d_{4}\sin(\lambda r)}
		\right)^{2}
		-d^{2}
		\right]^{b_{1}}
		[d_{1}\cos(\lambda t)+d_{2}\sin(\lambda t)]^{b_{2}}
		[d_{3}\cos(\lambda r)+d_{4}\sin(\lambda r)]^{b_{3}},
		\\
		\\
		\displaystyle
		p=\rho=
		\left[
		\left(
		\frac{d_{1}\cos(\lambda t)+d_{2}\sin(\lambda t)}
		{d_{3}\cos(\lambda r)+d_{4}\sin(\lambda r)}
		\right)^{2}
		-d^{2}
		\right]^{-b_{1}}
		\frac{(d_{1}^{2}+d_{2}^{2})(\beta_{2}+\beta_{3})}
		{
			[d_{1}\cos(\lambda t)+d_{2}\sin(\lambda t)]^{b_{2}+2}
			[d_{3}\cos(\lambda r)+d_{4}\sin(\lambda r)]^{b_{3}}
		},
	\end{array}
	\right.
	\label{45}
\end{equation}
\end{widetext}
where
\begin{eqnarray}
	b_{1}&=&\frac{b_{0}}{2}\left(\frac{d_{3}^{2}+d_{4}^{2}}{d_{1}^{2}+d_{2}^{2}}\right),
	\\
	b_{2}&=&\frac{\beta_{2}^{2}+\beta_{3}^{2}}{\lambda^{4}}
	-b_{0}\left(\frac{d_{3}^{2}+d_{4}^{2}}{d_{1}^{2}+d_{2}^{2}}\right)+1,
	\\
	b_{3}&=&\frac{\beta_{2}^{2}+\beta_{3}^{2}}{\lambda^{4}}
	+b_{0}\left(\frac{d_{3}^{2}+d_{4}^{2}}{d_{1}^{2}+d_{2}^{2}}\right)-1,
	\\
	d^{2}&=&\frac{d_{1}^{2}+d_{2}^{2}}{d_{3}^{2}+d_{4}^{2}}.
\end{eqnarray}

\paragraph{Particular solution for separable $F_0$.} As in the case $\delta=1$, another class of solutions can be obtained by assuming that the metric function $e^{2F_{0}}$ is separable in $t$ and $r$, i.e., $F_{0}(t,r)=A_{0}(t)+B_{0}(r)$.
Under this assumption Eq.~(\ref{EF1}) yields
\begin{widetext}
\begin{equation}
	e^{2F_{0}}=
	[d_{1}\cos(\lambda t)+d_{2}\sin(\lambda t)]^{\frac{2\beta_{0}}{\lambda^{2}}}
	[d_{3}\cos(\lambda r)+d_{4}\sin(\lambda r)]^{\frac{2}{\lambda^{2}}
		\left(
		\frac{\beta_{2}^{2}+\beta_{3}^{2}}{\lambda^{2}}-\beta_{0}
		\right)},
	\label{47}
\end{equation}
\end{widetext}
where the separation constant $\beta_{0}$ satisfies
\[
\beta_{0}=
-\frac{\beta_{2}^{2}+\beta_{3}^{2}+\beta_{2}\beta_{3}}
{\beta_{2}+\beta_{3}}.
\]

From Eq.~(\ref{EF2}) the energy-momentum profile becomes
\begin{equation}
	p=\rho=
	\frac{2\lambda^{2}(d_{1}^{2}+d_{2}^{2})
		[d_{1}\cos(\lambda t)+d_{2}\sin(\lambda t)]^{\frac{2(\beta_{2}^{2}+\beta_{3}^{2}+\beta_{2}\beta_{3})}{(\beta_{2}+\beta_{3})^{2}}}}
	{[d_{3}\cos(\lambda r)+d_{4}\sin(\lambda r)]^{\frac{2(2(\beta_{2}^{2}+\beta_{3}^{2})+\beta_{2}\beta_{3})}{(\beta_{2}+\beta_{3})^{2}}}}.
	\label{49}
\end{equation}

Therefore, we have obtained the general solution of the field equations for a stiff cylindrically symmetric fluid in the case $\delta=-1$. Additional particular solutions may be derived by selecting specific values of the separation constants.

\subsection{Specific case: $\delta =0$}

When $\delta =0$, Eq.~(\ref{15a}) admits the simple polynomial solution
\begin{equation}
	u=(at+b)(cr+d),
	\label{50}
\end{equation}
where $a$, $b$, $c$, and $d$ are constants of integration. This solution corresponds to the case in which the temporal and radial parts of $u$ are linear functions of $t$ and $r$, respectively.

Substituting Eq.~(\ref{50}) into the conservation equations
(\ref{11a}), and assuming the decomposition $F_i(t,r)=A_i(t)+B_i(r)$, the equations governing the metric functions $F_i$ reduce to
\begin{equation}
	\frac{1}{t}\frac{d}{dt}\left(t\dot{A}_i\right)
	-
	\frac{1}{r}\frac{d}{dr}\left(rB'_i\right)
	=\gamma_i ,
	\qquad i=2,3,
	\label{51}
\end{equation}
where $\gamma_i$ are separation constants.

\medskip

\textbf{Case I: $\gamma_i =0$.}

We first analyse the case in which the separation constants vanish. Under this assumption, Eqs.~(\ref{EF1})--(\ref{EF4}) lead to the following expressions for the metric functions
\begin{eqnarray}
	e^{2F_0}&=&(tr)^{m_2n_2+m_3n_3}\,
	\Pi\!\left(\ln\left|\frac{t}{r}\right|\right),\\
	e^{2F_i}&=&t^{2m_i}r^{2n_i},\qquad i=2,3,
	\label{53}
\end{eqnarray}
where, without loss of generality, we have chosen
$a=c=1$ and $b=d=0$ in Eq.~(\ref{50}). The constants
$m_i$ and $n_i$ are integration constants that satisfy
\[
m_2+m_3=1, \qquad n_2+n_3=1 .
\]
The function $\Pi$ depends only on the ratio $t/r$ and therefore describes the remaining freedom in the solution.

Substituting Eqs.~(\ref{53}) into Eq.~(\ref{EF3}) yields a differential equation for $\Pi$. Introducing the variable $x=\ln\left|t/r\right|$, the resulting equation can be written as
\begin{equation}
	(1-e^{2x})\frac{d^2\Pi}{dx^2}
	-2e^{2x}\frac{d\Pi}{dx}
	-2\left[(n_3-n_2)(m_3-n_3)+1\right]e^{2x}=0.
	\label{54}
\end{equation}
The general solution of Eq.~(\ref{54}) is
\begin{equation}
	\Pi(x)
	=
	(n_3-n_2)(m_3-n_3)x
	+
	\frac{a_0}{2}
	\ln\left|
	\frac{\sqrt{e^{2x}-1}}{e^{2x}}
	\right|,
\end{equation}
where $a_0\neq0$ is an integration constant.

Consequently, the metric component $e^{2F_0}$ takes the form
\begin{equation}
	e^{2F_0}
	=
	\frac{t^{2[m_3(n_3-n_2)+1-n_3^2]-a_0}}
	{r^{2n_2n_3}}
	\left(t^2-r^2\right)^{\frac{a_0}{2}},
	\label{55}
\end{equation}
while the corresponding energy density and pressure (recall that
$p=\rho$ for a stiff fluid) are given by
\begin{equation}
	p=\rho=
	\frac{\left[2(m_3n_3+1-m_3^2-n_3^2)-a_0\right]\,r^{2n_2n_3}}
	{t^{2[m_3(n_3-n_2)+2-n_3^2]-a_0}
		\left(t^2-r^2\right)^{\frac{a_0}{2}}}.
	\label{56}
\end{equation}

\medskip

A second class of solutions arises if we assume that
$F_0(t,r)$ is separable,
\[
F_0(t,r)=A_0(t)+B_0(r).
\]
Under this assumption the field equations yield
\begin{eqnarray}
	e^{2F_0}&=&t^{2\alpha_0}r^{2(m_2n_2+m_3n_3-\alpha_0)},\\
	e^{2F_i}&=&t^{2m_i}r^{2n_i}, \qquad i=2,3,
	\label{57}
\end{eqnarray}
and the energy-momentum profile becomes
\begin{equation}
	p=\rho=
	\frac{2(\alpha_0+m_2m_3)\,r^{2n_2n_3}}
	{t^{2[m_3(n_3-n_2)+2-n_3^2]}} .
	\label{59}
\end{equation}

The separation constant $\alpha_0$ is not independent but must satisfy
\[
\alpha_0=m_3(n_3-n_2)+1-n_3^2 .
\]

\medskip

\textbf{Case II: $\gamma_i\neq0$.}

We now consider the more general situation in which the separation
constants $\gamma_i$ do not vanish.
In this case the solutions of Eqs.~(\ref{EF1}) and (\ref{EF4}) take the form
\begin{widetext}
\begin{equation}
	e^{2F_0}
	=
	e^{z_1(at+b)^2(cr+d)^2+z_2(at+b)^2+z_3(cr+d)^2}
	[(at+b)(cr+d)]^{z_4}
	\Xi\!\left(\ln\left|\frac{at+b}{cr+d}\right|\right),
	\label{60}
\end{equation}

\begin{equation}
	e^{2F_i}
	=
	e^{\frac{\gamma_i(t+r^2)}{2}
		+\gamma_i\left(\frac{b}{a}t+\frac{d}{c}r\right)}
	(at+b)^{2\left(\frac{j_i}{a}-\gamma_i\frac{b^2}{2a^2}\right)}
	(cr+d)^{2\left(\frac{l_i}{c}-\gamma_i\frac{d^2}{2c^2}\right)},
	\qquad i=2,3 ,
	\label{61}
\end{equation}
\end{widetext}
where the constants satisfy
\[
\gamma_2+\gamma_3=0,
\qquad
j_2+j_3=a,
\qquad
l_2+l_3=c.
\]

For convenience we have introduced the parameters
\begin{eqnarray}
	z_1&=&\frac{\gamma_2^2+\gamma_3^2}{8a^2c^2}, \nonumber \\
	z_2&=&\frac{\gamma_2l_2+\gamma_3l_3}{2a^2c}-2d^2z_1, \nonumber\\
	z_3&=&\frac{\gamma_2j_2+\gamma_3j_3}{2ac^2}-2b^2z_1,\nonumber\\
    z_4&=&\frac{j_2l_2+j_3l_3}{ac}-2b^2d^2z_1-b^2z_2-d^2z_3 .\nonumber
\end{eqnarray}

Substituting the metric functions into Eq.~(\ref{EF2})
leads to a differential equation for the function $\Xi$.
Introducing the variable
\[
z=\ln\left|\frac{at+b}{cr+d}\right|,
\]
we obtain
\begin{equation}
	\left[e^{2z}-\left(\frac{a}{c}\right)^2\right]\frac{d^2\Xi}{dz^2}
	+2e^{2z}\frac{d\Xi}{dz}
	-2Ce^{2z}=0 ,
	\label{62}
\end{equation}
where $C$ is a constant determined by the parameters of the solution.

The general solution of Eq.~(\ref{62}) can be written as
\begin{equation}
	\Xi(z)
	=
	Cz
	+
	c_0\left(\frac{a}{c}\right)^2
	\ln\left|
	\frac{\sqrt{e^{2z}-\left(\frac{a}{c}\right)^2}}{e^{z}}
	\right|,
	\label{63}
\end{equation}
where $c_0$ is an arbitrary integration constant.

\medskip

These results complete the class of solutions corresponding to
$\delta=0$. As in the previous subsections, additional particular solutions may be obtained by imposing specific relations among the integration constants or by fixing the arbitrary functions appearing in the general expressions.

\section{Physical interpretation of the stiff-fluid solutions}
\label{SecIV}

In this section we investigate the physical properties of the classes of solutions obtained in the previous section for a cosmological perfect fluid obeying the stiff equation of state $p=\rho$. In particular, we analyze the geometrical and dynamical characteristics of the corresponding space-times, focusing on the kinematical properties of the fluid flow, the behaviour of the matter variables, the possible occurrence of curvature singularities, and the fulfillment of the standard energy conditions. 

Since the general solutions depend on the parameter $\delta$, which determines the functional form of the metric coefficient $u(t,r)$, the physical interpretation of the solutions naturally separates into three distinct cases, corresponding to $\delta=1$, $\delta=0$, and $\delta=-1$. These three values lead to qualitatively different functional structures of the metric functions and therefore to different dynamical behaviours of the corresponding cosmological models. In the following subsection we first provide a qualitative interpretation of these three classes of solutions.

\subsection{Interpretation of the three solution classes}

The families of solutions obtained in the previous section are classified according to the parameter $\delta$, which determines the mathematical structure of the auxiliary function $u(t,r)$ appearing in the metric coefficients. Because this function governs the temporal and radial dependence of the metric, the value of $\delta$ determines the global dynamical character of the corresponding space-time.

\paragraph{Case $\delta =1$.}

For $\delta=1$, the function $u(t,r)$ is expressed in terms of combinations of exponential functions of the form
\[
u(t,r)=\left(c_{1}e^{\lambda t}+c_{2}e^{-\lambda t}\right)
\left(c_{3}e^{\lambda r}+c_{4}e^{-\lambda r}\right),
\]
so that the metric coefficients exhibit exponential dependence on both the temporal and radial coordinates. As a consequence, the corresponding spacetimes generally display rapid dynamical evolution. 

In particular, the temporal behaviour of the metric functions may lead to models characterized by accelerated expansion or contraction, depending on the signs and magnitudes of the integration constants. In expanding configurations the exponential dependence may produce rapidly increasing scale factors along the spatial directions, while contracting solutions correspond to exponentially decreasing metric functions. Such behaviour is reminiscent of inflationary-type expansion in anisotropic geometries, although in the present case the symmetry is cylindrical rather than spatially homogeneous. 

The presence of arbitrary constants in the solutions allows the construction of a wide variety of dynamical configurations, ranging from models that approach asymptotic expansion to geometries that develop curvature singularities at finite values of the coordinates.\\

\paragraph{Case $\delta =0$.}

For $\delta=0$, the auxiliary function $u(t,r)$ becomes linear in both variables,
\[
u(t,r)=(at+b)(cr+d),
\]
which leads to metric coefficients that exhibit power–law dependence on the coordinates. In contrast with the exponential behaviour encountered in the previous case, these solutions correspond to more moderate dynamical evolution. 

The resulting spacetimes often display scaling properties typical of self-similar solutions in general relativity. Indeed, when the metric functions depend on powers of the coordinates, the geometry may admit homothetic symmetries or approximate scale invariance. Such solutions frequently arise in cosmological contexts describing intermediate stages of evolution between highly dynamical regimes.

Another characteristic feature of this class is the appearance of arbitrary functions, such as $\Pi(x)$ or $\Xi(z)$, which depend on the ratio of the temporal and radial variables. These functions represent a residual freedom in the integration of the field equations and allow the construction of families of models with different physical behaviours. Depending on the choice of these functions and the associated constants, the resulting spacetimes may describe expanding configurations, collapsing cylindrical matter distributions, or more complicated dynamical evolutions.\\

\paragraph{Case $\delta =-1$.}

When $\delta=-1$, the function $u(t,r)$ involves trigonometric functions of the coordinates,
\[
u(t,r)=\left[d_{1}\cos(\lambda t)+d_{2}\sin(\lambda t)\right]
\left[d_{3}\cos(\lambda r)+d_{4}\sin(\lambda r)\right].
\]
In this situation the metric functions exhibit oscillatory behaviour in both the temporal and radial directions. 

Such oscillatory dependence introduces periodic features in the gravitational field, which may lead to models in which the kinematical quantities of the fluid, such as the expansion scalar, change sign during the evolution. Physically, this may correspond to alternating phases of expansion and contraction of the cylindrical matter distribution. These solutions may therefore be interpreted as representing cyclic or wave-like gravitational configurations supported by a stiff fluid source.

The oscillatory structure also implies that the geometry may remain bounded within certain coordinate ranges, although curvature singularities can still arise for particular choices of the integration constants. Consequently, this class of solutions provides interesting examples of dynamical cylindrical space-times with periodic behaviour in the metric functions.

\subsection{Kinematical quantities of the fluid flow}

We now examine the kinematical properties of the cosmological fluid associated with the class of cylindrically symmetric solutions obtained in the previous sections. These properties provide important information regarding the dynamical evolution of the spacetime and the behaviour of the matter distribution.

Since the metric has been constructed in comoving coordinates, the fluid four-velocity is aligned with the temporal coordinate direction. It can therefore be written as $u^{\mu}=e^{-F_{0}}\delta^{\mu}_{0}$, which satisfies the normalization condition $u^{\mu}u_{\mu}=1$. 
The kinematical behaviour of the fluid flow is characterized by the expansion scalar $\Theta$, the shear tensor $\sigma_{\mu\nu}$, and the four-acceleration $\dot{u}_{\mu}$. These quantities are defined respectively by
\begin{equation}
	\Theta=u^{\mu}{}_{;\mu}, 
	\sigma_{\mu\nu}=u_{(\mu;\nu)}-\frac{1}{3}\Theta h_{\mu\nu}-\dot{u}_{(\mu}u_{\nu)}, 
	\dot{u}_{\mu}=u_{\mu;\nu}u^{\nu},
\end{equation}
where  $h_{\mu\nu}=g_{\mu\nu}+u_{\mu}u_{\nu}$ is the projection tensor onto the three–space orthogonal to the fluid flow.\\


\paragraph{Expansion scalar.}
The expansion scalar measures the rate of change of the volume of a fluid element as it evolves along the flow lines. A direct computation for the present metric yields
\begin{equation}
	\Theta=e^{-F_{0}}\left(\dot{F}_{1}+\dot{F}_{2}+\dot{F}_{3}\right),
\end{equation}
where an overdot denotes differentiation with respect to the time coordinate $t$. The sign of $\Theta$ determines whether the cosmological configuration undergoes expansion or contraction. In particular, $\Theta>0$ corresponds to an expanding spacetime, whereas $\Theta<0$ indicates a contracting configuration.

In particular, for the power-law solutions obtained in the case $\delta=0$, where the metric functions take the form
$ F_{i}=m_{i}\ln t+n_{i}\ln r$, the expansion scalar becomes
\begin{equation}
	\Theta=e^{-F_{0}}\frac{m_{1}+m_{2}+m_{3}}{t}.
\end{equation}
This expression shows that the expansion rate decreases with time as $t^{-1}$, which is characteristic of many cosmological solutions exhibiting self–similar behaviour. Using the constraint relations among the parameters $m_i$ derived from the Einstein field equations, this quantity can be expressed entirely in terms of the independent parameters of the model.\\


\paragraph{Shear tensor.}
The shear tensor $\sigma_{\mu\nu}$ describes the anisotropic deformation of the fluid flow, representing distortions in the shape of fluid elements without change in their volume. The magnitude of this anisotropic deformation is characterized by the shear scalar
$\sigma^{2}=\frac{1}{2}\sigma_{\mu\nu}\sigma^{\mu\nu}$.
For the class of metrics considered here, a straightforward calculation leads to
\begin{equation}
	\sigma^{2}=\frac{e^{-2F_{0}}}{3}
	\left[
	\dot{F}_{1}^{2}+\dot{F}_{2}^{2}+\dot{F}_{3}^{2}
	-\dot{F}_{1}\dot{F}_{2}
	-\dot{F}_{2}\dot{F}_{3}
	-\dot{F}_{3}\dot{F}_{1}
	\right].
\end{equation}

In general, the solutions obtained in this work possess non-vanishing shear, reflecting the anisotropic character of cylindrically symmetric spacetimes. The expansion rates along the radial, azimuthal, and axial directions are governed by the time derivatives of the functions $F_{1}$, $F_{2}$, and $F_{3}$, which need not coincide. Consequently, the evolution of the spacetime typically proceeds in an anisotropic manner.\\


\paragraph{Isotropization.}
An important issue in anisotropic cosmological models concerns the possibility of isotropization during the cosmic evolution. This behaviour can be investigated by considering the ratio of the shear scalar to the expansion scalar,
$ \sigma/\Theta$.
If this ratio tends to zero as $t\rightarrow\infty$, the anisotropic distortions of the fluid flow become negligible compared with the overall expansion rate, and the spacetime approaches isotropic behaviour at late times.

For the present class of solutions the asymptotic behaviour of $\sigma/\Theta$ depends sensitively on the values of the integration constants that characterize the metric functions. In many cases the shear decreases at a rate comparable to that of the expansion, implying that the ratio $\sigma/\Theta$ approaches a constant value and the spacetime remains anisotropic throughout its evolution. However, for particular choices of parameters the shear may decay more rapidly than the expansion, allowing the models to approach isotropy at sufficiently late times.

In the specific case of the power-law solutions discussed above, the shear scalar typically behaves as $\sigma^{2}\propto t^{-2}$, while the expansion scalar varies as $\Theta\propto t^{-1}$. Although both quantities decrease during the evolution, their ratio does not necessarily vanish asymptotically, indicating that isotropization is not guaranteed for arbitrary values of the integration constants.

\subsection{Energy density and energy conditions}

The matter source considered in the present work is a perfect fluid obeying the stiff equation of state $ p=\rho$.
This equation of state corresponds to the limiting case of a relativistic fluid in which the speed of sound equals the speed of light, $c_s^2=dp/d\rho=1$. Stiff fluids arise naturally in several contexts in relativistic cosmology, including models of the very early universe and in the effective description of a massless scalar field. Because of their extreme rigidity, such fluids provide a useful idealization for studying the dynamical behaviour of highly relativistic matter distributions in strong gravitational fields.

For the class of cylindrically symmetric solutions obtained in the previous sections, the energy density $\rho$ is determined directly from the Einstein field equations and depends on the metric functions $F_i(t,r)$ that characterize the spacetime geometry. The explicit functional form of $\rho$ therefore varies for the different families of solutions corresponding to the values $\delta=1,0,-1$.

As an illustrative example, in the case $\delta=0$ the metric functions take a power–law form and the energy density becomes
\begin{equation}
	\rho =
	\frac{\left[ 2\left( m_{3}n_{3}+1-m_{3}^{2}-n_{3}^{2}\right)-a_{0}\right]
		\, r^{2n_{2}n_{3}}}
	{t^{2\left[ m_{3}\left( n_{3}-n_{2}\right)+2-n_{3}^{2}\right]-a_{0}}
		\left( t^{2}-r^{2}\right) ^{\frac{a_{0}}{2}}}.
\end{equation}
This expression shows that the density depends on both the temporal and radial coordinates, reflecting the inhomogeneous and anisotropic nature of the cylindrically symmetric configurations considered here. The behaviour of the density during the evolution of the system is governed by the integration constants appearing in the solution, which determine whether the density increases or decreases with time and radial distance.

For a physically meaningful cosmological model the energy density must be non–negative, $\rho \geq 0$. 
This requirement imposes constraints on the integration constants entering the solutions, ensuring that the matter distribution remains physically admissible throughout the spacetime region under consideration.

	It is also instructive to examine the standard energy conditions associated with the energy-momentum tensor of the fluid.
	For a perfect fluid in a four-dimensional spacetime, these conditions take a form that depends only on the components of $T_{\mu\nu}$ and not on the specific symmetry of the geometry~\cite{Hawking:1973uf,Brassel:2021ent}.
	Consequently, the pointwise energy conditions for a perfect fluid in a cylindrically symmetric spacetime are formally identical to those employed in spherical symmetry: they are expressed entirely in terms of the energy density $\rho$ and the pressure $p$, without reference to the underlying metric symmetry.

For a perfect fluid these conditions can be written as follows:
\begin{itemize}
	\item Weak energy condition (WEC): $\rho \geq 0$ and $\rho+p \geq 0$,
	\item Strong energy condition (SEC): $\rho+p \geq 0$ and $\rho+3p \geq 0$,
	\item Dominant energy condition (DEC): $\rho \geq |p|$.
\end{itemize}

	The key difference between spherical and cylindrical symmetry lies not in the form of the energy conditions themselves but in the structure of the Einstein field equations that relate $T_{\mu\nu}$ to the metric functions.
	In spherical symmetry the field equations impose different relationships among the metric coefficients, which in turn constrain the admissible matter profiles in a manner that is specific to that symmetry class.
	In the present cylindrically symmetric case the field equations~\eqref{EF1}--\eqref{EF4} couple the energy density and pressure to the metric functions through the auxiliary variables $h_i$, $k_i$, and their combinations, leading to constraints that are structurally distinct from those arising in spherical symmetry.
	Nevertheless, the pointwise energy conditions written above remain the same, and their verification proceeds directly from the expressions for $\rho$ and $p$.

For the stiff fluid equation of state $p=\rho$, these conditions simplify considerably. Thus, the weak, strong, and dominant energy conditions are all satisfied provided that
$\rho \geq 0$. 
Consequently, once the positivity of the energy density is ensured, all standard energy conditions are automatically fulfilled by the matter source. This feature reflects the physically reasonable character of stiff matter, which respects the causal propagation of energy and momentum and does not violate the usual requirements imposed in classical general relativity.

Therefore, the admissibility of the solutions obtained in this work reduces essentially to identifying the ranges of the integration constants for which the energy density remains non-negative and finite in the spacetime region of interest.

\subsection{Curvature invariants and singularities}

A fundamental aspect in the analysis of any exact solution of the Einstein field equations is the investigation of the possible occurrence of spacetime
singularities. In general relativity, genuine physical singularities are characterized by the divergence of curvature invariants constructed from the Riemann tensor, which signal the breakdown of the classical geometrical description of spacetime. The study of such invariants therefore provides a useful and coordinate–independent method for determining whether a given solution possesses pathological regions where the gravitational field becomes unbounded.

For the class of cylindrically symmetric spacetimes considered in this work, the matter source is a stiff perfect fluid satisfying the equation of state $p=\rho$. Thus, the trace of the Einstein field equations provide the following relation for the Ricci scalar: $R=2\rho$.
It follows that whenever the energy density diverges the curvature of spacetime also diverges, indicating the presence of a curvature singularity.

In addition to the Ricci scalar, an important invariant characterizing the strength of the gravitational field is the Kretschmann scalar, 
$K=R_{\alpha\beta\gamma\delta}R^{\alpha\beta\gamma\delta}$.
This invariant contains quadratic combinations of the components of the Riemann tensor and therefore provides a more complete measure of the curvature of spacetime. In the present case, where the metric coefficients depend on the functions $F_i(t,r)$, the Kretschmann scalar involves quadratic combinations of their derivatives with respect to both the temporal and radial coordinates.
Although the explicit expression for $K$ is rather lengthy, its general
structure can be understood schematically as
\begin{equation}
	K \sim
	(\dot{F}_{i})^{4}
	+(F_{i}^{\prime})^{4}
	+(\dot{F}_{i}F_{j}^{\prime})^{2}
	+\ddot{F}_{i}^{2}
	+F_{i}^{\prime\prime 2},
\end{equation}
where overdots and primes denote derivatives with respect to $t$ and $r$, respectively, and repeated indices $i,j=0,1,2,3$ are summed over. This expression shows that the Kretschmann scalar depends on quadratic combinations of both first and second derivatives of the metric functions. Therefore, curvature singularities generally occur whenever these derivatives diverge or when the metric functions themselves develop pathological behaviour.

The occurrence of singularities depends on the particular family of solutions characterized by the parameter $\delta$. In the power-law solutions corresponding to $\delta=0$, the metric functions typically contain logarithmic dependencies on $t$ and $r$, which lead to divergences of the curvature invariants as $t\rightarrow 0$. This behaviour is reminiscent of the initial singularity commonly encountered in anisotropic cosmological models and may be interpreted as describing an early-time phase where the curvature of spacetime becomes arbitrarily large.

For the class of solutions with $\delta=1$, the metric functions involve exponential dependencies on the temporal and radial coordinates. In these models the curvature invariants may either grow or decay exponentially, depending on the values of the integration constants. In particular, certain choices of parameters may lead to rapidly increasing curvature, signalling the development of strong gravitational fields at large times or large radial distances.

Finally, for $\delta=-1$ the metric functions exhibit trigonometric dependencies, leading to oscillatory behaviour of the gravitational field. Although the curvature invariants remain finite for many parameter choices, singularities may still arise at specific hypersurfaces where the metric functions vanish or where their derivatives diverge. Such singular behaviour is associated with the nodal structure of the trigonometric solutions and reflects the highly dynamical character of the corresponding spacetime configurations.

\section{Conclusion}\label{SecV}

In this work we have obtained the general solutions of the Einstein field equations for a cylindrically symmetric spacetime filled with a stiff perfect fluid obeying the equation of state $p=\rho$. By adopting the Marder form of the metric and applying a systematic separation of variables procedure, the field equations were solved in full generality, leading to three distinct classes of exact solutions characterized by the parameter $\delta = 1,0,-1$. These three cases correspond respectively to exponential, power-law, and trigonometric structures of the metric functions, thereby providing a unified framework that encompasses a wide range of possible dynamical behaviours within cylindrically symmetric geometries.

The physical analysis of these solutions shows that the corresponding cosmological models generally exhibit anisotropic expansion, as reflected by the non-vanishing shear tensor associated with the fluid flow. The kinematical quantities, including the expansion scalar and the shear scalar, are determined by the integration constants that arise in the solutions and govern the dynamical evolution of the spacetime. The energy density obtained from the Einstein equations depends explicitly on the temporal and radial coordinates, illustrating the intrinsically inhomogeneous character of the configurations. Physical viability requires the positivity of the energy density, which simultaneously guarantees the validity of the weak, dominant, and strong energy conditions for the stiff fluid. The investigation of curvature invariants, such as the Ricci and Kretschmann scalars, further reveals the possible presence of curvature singularities associated with divergences of the energy density or of the derivatives of the metric functions.

Beyond the specific properties of the solutions, the results presented here highlight the richness of the gravitational dynamics generated by stiff matter in cylindrically symmetric spacetimes. The exponential solutions may describe rapidly evolving regimes of the gravitational field, the power-law solutions are closely related to self-similar cosmological behaviour, and the trigonometric solutions provide examples of oscillatory gravitational dynamics. Together, these families illustrate how a simple relativistic equation of state can generate a broad spectrum of geometrical and dynamical structures in exact solutions of the Einstein equations.

The significance of these solutions extends beyond the immediate context of perfect-fluid cosmologies. Since a stiff fluid is dynamically equivalent to a massless scalar field, the models obtained here may also be interpreted as exact scalar-field configurations with cylindrical symmetry, which are of particular interest in studies of the early universe and anisotropic cosmological dynamics. Moreover, exact solutions of this type provide valuable analytical laboratories for investigating the interplay between geometry and matter in general relativity, and they can serve as useful benchmarks for numerical studies of cylindrical gravitational systems.

	Although the solutions obtained in this work are idealized configurations, they may capture qualitative features relevant to the interpretation of certain cosmological observations.
	The anisotropic and inhomogeneous character of these cylindrically symmetric models reflects, in a simplified setting, the type of deviations from perfect isotropy that are constrained by modern CMB experiments.
	In particular, the Planck satellite has placed increasingly stringent upper bounds on anisotropic expansion and on global shear at the epoch of recombination~\cite{Planck:2019evm,Saadeh:2016sak}.
	While the spacetimes constructed here are not directly applicable as realistic cosmological models, the presence of non-vanishing shear and the possibility of late-time isotropization for certain parameter ranges illustrate the kind of anisotropic dynamics that could have been operative in the very early Universe and that may leave subtle imprints on observable quantities.
	Furthermore, cylindrical geometries provide a tractable analytical framework for exploring the evolution of primordial inhomogeneities and the gravitational collapse of extended objects, topics that remain central to the interpretation of large-scale structure surveys and of the stochastic gravitational-wave background~\cite{Caprini:2018mtu}.
	In this sense, the families of exact stiff-fluid solutions derived here can serve as a theoretical laboratory for testing ideas about the transition from an anisotropic early phase to the highly isotropic Universe we observe today.

The framework developed in this work can also be extended in several directions. Possible generalizations include the incorporation of additional physical fields such as electromagnetic sources, dissipative or viscous fluids, or scalar fields with nontrivial potentials. It would also be of interest to explore analogous configurations in modified theories of gravity, including $f(R)$ and scalar-tensor models, where stiff matter often plays a significant dynamical role. Furthermore, the solutions presented here may offer useful insights into the study of anisotropic gravitational collapse, cylindrical gravitational waves, and other phenomena associated with strongly inhomogeneous spacetime geometries.

In this sense, the present analysis contributes to the continuing effort to enlarge the catalogue of exact solutions of the Einstein field equations and to deepen our understanding of anisotropic cosmological dynamics. The families of stiff-fluid solutions obtained here provide a versatile set of models that may serve as a starting point for further investigations into the structure and evolution of cylindrically symmetric spacetimes in relativistic gravitation.

\acknowledgments{We would like to thank the anonymous reviewer for comments and suggestions that helped us to improve our manuscript. FSNL acknowledges support from the Funda\c{c}\~{a}o para a Ci\^{e}ncia e a Tecnologia (FCT) Scientific Employment Stimulus contract with reference CEECINST/00032/2018, and funding through the grant UID/04434/2025.}



\end{document}